# Improving Security Levels of IEEE 802.16e Authentication By Diffie-Hellman Method

Mohammad Zabihi[1], Ramin Shaghaghi[2], Mohammad Esmail kalantari[3]

[1] Department of Electrical Engineering, Islamic Azad University, Shahre- Rey Branch, Tehran, Iran

[2] Department of Electrical Engineering, Islamic Azad University, Shahre- Rey Branch, Tehran, Iran

[3] Department of Electrical Engineering, Islamic Azad University, Shahre- Rey Branch, Tehran, Iran

**Abstract**

In this paper, we proposed an authentication method according to Diffie-Hellman. First, we introduce different methods for authentication in IEEE.802.16 then we proposed an authentication method according to Diffie-Hellman and in the last we compare different methods for authentication to improve security in IEEE802.16e. CPN is a useful for simulation and compare protocol together so we use CPN tools in this paper..

## 1. Introduction

The importance of IEEE 802.16, Worldwide Interoperability for Microwave Access (Wimax) is growing and will complete with technologies such as 3G. The acceptance and adoption of technologies also depend on security. IEEE 802.16 Wimax standard consists of a protocol stack with well-defined interfaces. The Wimax protocol layer contains MAC layer and PHY layer. MAC layer includes three sub-layers contain of: The Service Specific Convergence Sub-layer (MAC CS), the MAC Common Part Sub-layer (MAC CPS) and the Security Sub-layer or Privacy Sub-layer. The former IEEE 802.16 standards used the Privacy and Key Management (PKM) protocol which had many critical drawbacks. In IEEE 802.16e, a new version of this protocol called PKMv2 is released. The authentication and key management protocols are specified in the security sub layer of IEEE 802.16 standard. The security sub layer is meant to provide subscribers with privacy and authentication and operators with strong protection from theft of service. Authentication options are: unilateral authentication, mutual authentication and no authentication sections IEEE S02.16e-2005 standard states that PKM has two versions PKMvl and PKMv2, and it allows for four types of authentication":

- A.RSA base authentication-PKI system(public key infrastructure).
- EAP based authentication (optional).
- RSA based authentication followed by EAP authentication.
- Diffie Hellman base authentication.

## 2. RSA Base authentication protocol

2.1 Pkmv1 Authentication Protocol

SS uses Message 1, formally named as the Authentication Information Message, to push its X.509 certificate which identifies its manufacturer to BS. BS uses this certificate to decide whether SS is a trusted device. BS may use this message in order to allow access only to devices from recognized manufacturers, according to its security policy. SS sends Message 2, named as the Authorization Request immediately after Message 1(figure.1). Message 2 consists of SS's X.509 certificate with the SS public key, its security capabilities which are actually the authentication and encryption algorithms that SS support, and the security association identity (SAID) which is the id of the secure link between SS and BS. Using the certificate, BS determines whether to authorize SS; and the public key of SS which is also in the certificate lets BS construct Message 3 [1]. If successful, namely SS is authorized after BS verifies its certificate, BS responds with Message 3, the Authorization Reply. This message includes the AK, encrypted using the RSA public-key encryption protocol using the public-key of SS which was obtained in the previous message, the lifetime of the AK as a 32-bit

164

unsigned number in unit of seconds, the sequence number for AK as a 4-bit value and the list of SA descriptors each including an SAID and the SA cipher suit [1].

the end of the RSA authorization exchange, both SS and BS are authenticated by each other [5].

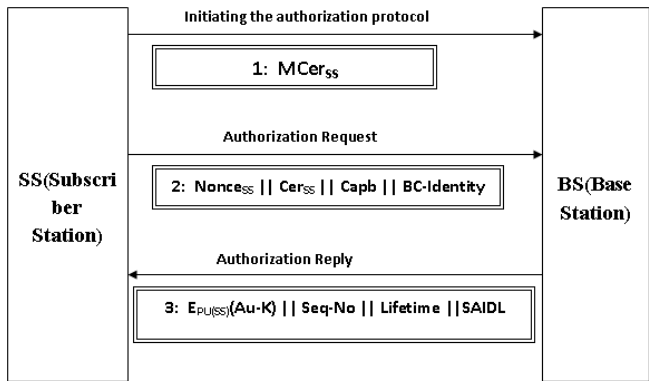

Fig. 1 pkmv1 authentication protocol.

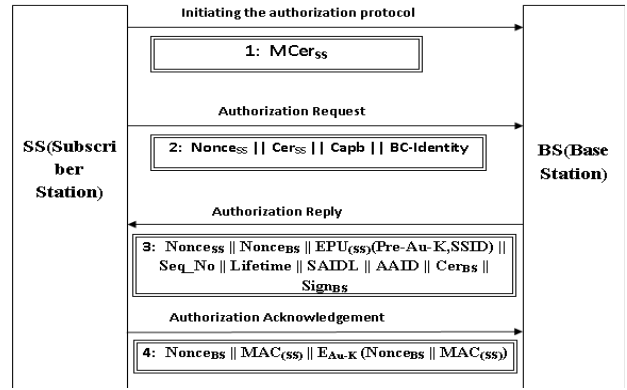

Fig. 2 pkmv2 authentication protocol.

## 2.2 Pkmv2 Authentication Protocol

The latest standard, IEEE 802.16e-2005, includes a new version (PKMv2) of the protocol that caters for the shortcomings of the first version. PKMv2 supports two different mechanisms for authentication: the SS and the BS may use RSA-based authentication or Extensible Authentication Protocol (EAP) -based authentication. We will focus in this paper on RSA based authentication for PKMv2 authentication protocol. The flow of messages exchange in RSA-based authentication is shown as follows(figure.2): The SS initiates the RSA-based mutual authentication process by sending two messages. The first message contains the manufacturer X.509 certificate. The second, authorization request message, contains the SS's X.509 certificate, 64-bit SS random number Ns, list of security capabilities that the SS supports, the SAID and the SS signature. If the SS is authenticated and authorized to join the network, the BS sends an authorization reply message. In the response message, the BS includes the 64-bit SS random number Ns received, its own 64-bit random number Nb, a 256-bit key pre-primary authorization key (pre-PAK) encrypted with the SS's public key, the pre-PAK key lifetime and its sequence number, a list of SAIDs (one or more), the BS's X.509 certificate and BS's signature in the authorization reply. The SS verifies liveness by comparing the Ns it sent with the received Ns in the authorization response message. It then extracts the PAK, because only the authorized SS can extract the PAK. This can be used as a proof of authorization. Finally, the last message of this authentication is send by the SS to confirm the authentication of the BS. The SS includes the BS random number Nb received in the authorization response message, used to proof liveness, the SS's MAC address and a cryptographic checksum of the message. At the Extensible Authentication Protocol (EAP) is an authentication framework that is widely used in WiFi/802.11 and Wimax/ 802.16 wireless networks. EAP is a basis to transfer authentication information between a client and a network. It provides a basic request/response protocol framework over which to implement a specific authentication algorithm, so called EAP method. Commonly used EAP methods are EAP-MD5, EAPLEAP, EAP-TLS, EAP-TTLS and EAP-PEAP. Within the EAP framework, three entities are involved in the authentication process: Supplicant, Authenticator, and Authentication Server. The supplicant is a user that is trying to access the network. It is also known as the peer. The authenticator is an access point (AP) that is requiring EAP authentication prior to granting access to a network. It provides users a point of entry into the network. The authentication server (AS) is the entity that negotiates the use of a specific EAP method with an EAP supplicant, then validates the supplicant, and authorizes access to the network.

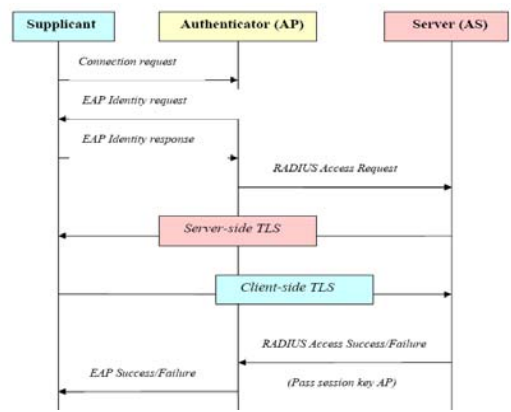

Fig. 3 EAP TLS base authentication.

Typically, the supplicant is a mobile station (MS) and the authentication server is a Remote Authentication Dial-In User Service (RADIUS.Figure.3 shows the brief message flow of EAP-TLS in a WLAN network. The AP creates a RADIUS Access Request using the supplicant's identity and sends it to the AS.The AS then provides its certificate to the supplicant and asks for the supplicant's certificate. The supplicant provides its certificate to AS if the received AS's certificate is valid. After the AS validates the supplicant's certificate, it will send the result message RADIUS Access Success/Failure to deny or permit access to the network [3].

## 4. Proposed Protocol

Diffie-Hellman key exchange (D-H) is a cryptographic protocol that allows two parties that have no prior knowledge of each other to establish together a shared secret key over an insecure communications channel. Then they use this key to encrypt subsequent communications using a symmetric-key cipher. The scheme was first published publicly by Whitfield Diffie and Martin Hellman in 1976. the Diffie- Hellman exchange by itself does not provide authentication of the communicating parties and is thus susceptible to a man-in-the-middle attack. An attacking person in the middle may establish two different Diffie-Hellman key exchanges, with the two members of the party "A" and "B", appearing as "A" to "B", and vice versa, allowing the attacker to decrypt (and read or store) then re-encrypt the messages passed between them. A method to authenticate the communicating parties to each other is generally needed to prevent this type of attack [2]. As shown in Figure.4, AS sends a request message to the BS that includes the certificate. Then the AS responds this message by sending (Cert BS,P(nonce),H(Ybs||f(nonce)) to A. Being a RSA encryption, P can encrypt nonce and User A can decrypt to receive nonce.

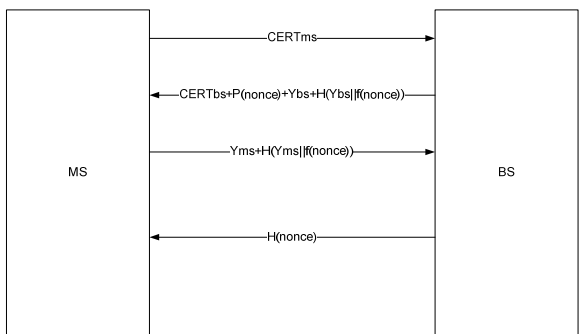

Fig. 4 proposed protocol.

Now, nonce is shared by MS and BS. And then, AS sends (Yms, H($Y_{ms}$ ||nonce)) to BS, BS calculates H '($Y_{AS}$ ||nonce). If H '($Y_{AS}$ ||nonce) is equal to H($Y_{AS}$ ||nonce), BS believes this message sent by AS, or interrupts this communication. Similarly, AS calculates H '($Y_{BS}$ ||f(nonce)) by $Y_{BS}$ and from AS. If H '($Y_{BS}$ ||f(nonce)) is equal to H ($Y_{BS}$ ||f(nonce)), AS believes this message sent by BS and calculates the K= $(Y_A)X_B$ mod q, or interrupts this communication. After A sends a massage H(nonce) as a confirmation signal to BS, AS and BS calculate the:

$$AK= (Y_{AS})X_{BS} \bmod q=(Y_{BS})X_{AS} \bmod q \quad (1)$$

## 5. Threat analysis

intercepts messages during the process of communication establishment or a public key exchange and then retransmits them, tampering the information contained in the messages, so that the two original parties still appear to be communicating with each other. In a man-in-the-middle attacks, the intruder uses a program that appears to be the (access point) AP to SS and appears to be the SS to AP. Denial of Service (DOS) attack is an incident in which a subscriber is deprived of the service of a resource they would normally expect to have. A considerable amount of denial of service attacks implement across the Internet by flooding the propagation medium with noise and forge messages. The victim is overwhelmed by the sheer volume of traffic, with either its network bandwidth or its computing power exhausted by the flood of information. Almost all the DOS vulnerabilities in Mobile wimax standard are due to unauthenticated or unencrypted management messages. We discussed these vulnerabilities in three processes: the initial network process, resource saving process and handover process[7]. eavesdropping of management messages is a critical threat for users and a major threat to a system. For example, an attacker could use this vulnerability to verify the presence of a victim at its location before perpetrating a crime. Additionally, it might be used by a competitor to map the network. Another major vulnerability is the encryption mode based on data encryption standard (DES). The 56 bit DES key is easily broken with modern computers by brute force attack. Furthermore, the DES encryption mode includes no message integrity or replay protection functionality and is thus vulnerable to active or replay attacks. The secure AES encryption mode should be preferred over DES. Eavesdropping mostly affects the transfer of information and rarely causes system outage. The assessment of the eavesdropping threat is minor to the system but high for the user [8] .in table.1 we show vulnerability for each method. the proposed authentication protocol has best function in comparison with other protocol.

## 6. Protocol Analysis by CPN

We used a Petri net to model our security protocol. Colored Petri Nets (CP-nets or CPNs) is a graphical We used a Petri net to model our security protocol.
Colored Petri Nets (CP-nets or CPNs) is a graphical language for constructing models of concurrent systems and analysing their properties. CPnets is a discrete-event modeling language combining Petri nets and the functional programming language CPN ML which is based on Standard ML . A CPN model of a system describes the states of the system and the events (transitions) that can cause the system to change state. By making simulations of the CPN model, it is possible to investigate different scenarios and explore the behaviors of the system. Very often, the goal of simulation is to debug and investigate the system design. CP-nets can be simulated interactively or automatically[11]. Petri nets are composed from graphical symbols designating places (shown as circles), transitions (shown as rectangles), and directed arcs (shown as arrows). The Petri net model is illustrated in Figure 7. The model is simulated with the time color Petri net simulation tool. The basic information about the size of the state space and standard behavioral properties of the CPN model can be found in the state space report. For the CPN model of the proposed protocol in this study, the state space report is shown in Figure 6. As shown in Figure 6, we have data about "State Space statistics (Strongly-connected-component/Scc graph)", "Liveness Properties (Dead Markings, Dead Transition Instances, and Live Transition Instances)". The state space statistics inform about the size of the state space. For the model of proposed protocol, there are 11 nodes and 10 arcs. If the nodes and arcs in the state space and Scc graph are equal, it means that there are no cycles in the model. The number of nodes and arcs in the state space and Scc graph of two protocols are equal. It means that the token will not fall in a loop, and we have finite-occurrence sequences. A dead marking is a mark in which no element is enabled. This means that the marking corresponding to node 11 in Figure 7 is a dead marking. A transition is live if from any reachable marking we can always find an occurrence sequence containing the transition. As shown in Figure 6, there are no live transitions[10].Two protocols have a dead marking. So, they have not "live transitions". It is noted that no transition could be enabled from the dead marking. Also, there are no dead transitions in two protocols. A transition is dead, if there is no reachable marking in which it is enabled.There is no dead transition, which means that each transition in the protocol has the possibility to occur at least once. If a model has a dead transition, then it

```
            Statistics
            ----------

            State Space
   Nodes:   11
    Arcs:   10
    Secs:    0
   Status: Full

            Scc Graph
   Nodes:   11
    Arcs:   10
    Secs:    0
```

Table 1: Attack in authentication protocol.

| attack | PKMv2(with nonce) | EAP | Proposed Protocol |
|---|---|---|---|
| MITM | About weak | resistant | resistant |
| Replay | About weak | resistant | resistant |
| Interception | resistant | resistant | resistant |

Fig. 6 state space report.

corresponds to parts of the model that can never be activated. Hence, we can remove dead transitions from the model without changing the behavior of it. To compare the service delivery time of four protocols, we assume that each transition in two protocols takes 5 time units [10]. Several different kinds of output can be generated for data collector monitors. All of the data that is collected by a data collector can be saved in a data collector log file (Table.2).The log file also contains information about the steps and model times at which the data was collected [11].

Table 2: Timed static for propose protocol.

| Name | Count | Avrg | 90% Half Length | 95% Half Length | 99% Half Length | SSD | Variance | StD |
|---|---|---|---|---|---|---|---|---|
| Marking_size_simple'Data_Received_1 | 4 | 1.000000 | 0.000000 | 0.000000 | 0.000000 | 0.000000 | 0.000000 | 0.000000 |
| Marking_size_simple'b_1 | 6 | 0.200000 | 0.332389 | 0.424105 | 0.665108 | 8.000000 | 0.163265 | 0.404061 |
| Marking_size_simple'recive_ms_1 | 4 | 0.700000 | 0.760976 | 1.029080 | 1.889018 | 20.500000 | 0.418367 | 0.646813 |

Simulation steps executed: 10
Model time: 50.0

## 7. Conclusion

In this paper, we have introduced a authentication protocol for Wimax network. The proposed protocol provides mutual authentication, key exchange between MS and BS, so the probability of some threats such as eavesdropping, MITM and replay is reduced. Also, we have compared the performance of proposed protocol with another protocols by using CPN Tools. It has been shown that by omission of some transactions, the number of place and transition are reduced. Also, the execution time is reduced significantly, and the network resources are reserved. The state space reports are shown in Table 3.

Table 3: state space statistic.

| property | Pkmv2(RSA) | EAP | proposed |
|---|---|---|---|
| Simulation step | 7 | 13 | 10 |
| Model time | 40 | 90 | 50 |
| Number of place | 12 | 12 | 10 |
| Number of transition | 5 | 9 | 5 |
| State space nodes | 10 | 19 | 11 |
| State space arc | 9 | 22 | 10 |
| Dead marking | 20,21,22 | 6,19 | 11 |

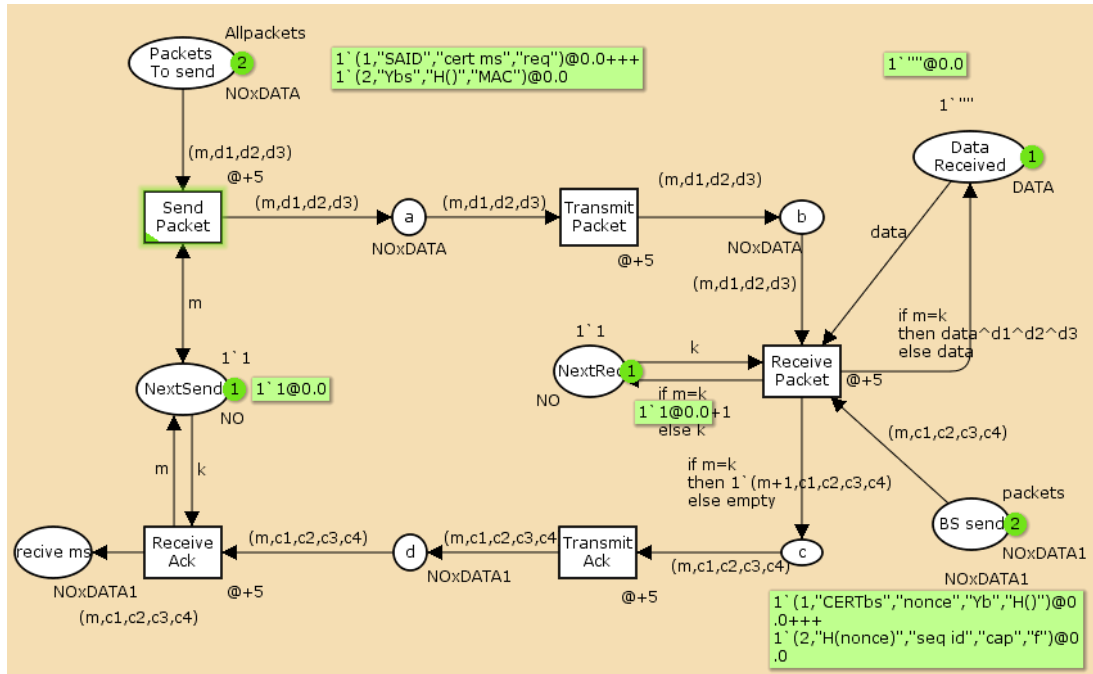

Fig. 7 petri net model for proposed protocol.

## References

[1] E.Yuksel," Analysis of the PKMv2 Protocol in IEEE 802.16e-2005 Using Static Analysis", Technical University of Denmark,Feb 2007.
[2] N.Li," Research on Diffie-Hellman Key Exchange Protocol",IEEE,2010.
[3] L.Han,"a threat analysis of the Extensible Authentication Protocol",Apr 2006.
[4] J.Hur,H.Shim,P.Kim,H.Yun,N.Oak song,"security considerations for handover shemes in mobile wimax networks",IEEE,2008.
[5] A.Taha,A.Abdel hamid,A.Tahar,"formal verification of IEEE 802.16 security sublayer using scyther tools", ESRGroups France,2009
[6] Z.You,X.Xie,W.Zheng," Verification and Research of a Wimax authentication protocol Based on SSM",ICETC,2010.
[7] C. J. Kaufman, Rocky Mountain Research Lab., Boulder, CO, private communication, May 1995.
[8] M. Bogdanoski, P.Latkoski, A.Risteski, B.Popovski," IEEE 802.16 Security Issues: A Survey",Telecommunication forum,2008.
[9] H.Tseng, R.Hong Jan, W.Yang,"A chaotic maps-base key agreement protocol that preseres user anonymity",IEEE ICC,2009.
[10] M.Shaikhan,A.Sobhani,M.E.Kalantari," Modification of Mobile Web Shopping Protocol Using GAA and Analysis by Colored Petri Nets",SATCCN,2011.
[11] K. Jensen , L.Michael Kristensen, L. Wells," Coloured Petri Nets and CPN Tools for Modelling and Validation of Concurrent Systems" ,Department of Computer Science,2008.
[12] J.Huang, C.Tser Huang," Secure Mutual Authentication Protocols for Mobile Multi-hop Relay WiMAX Networks against Rogue Base/Relay Stations"IEEE,2011.

[13] S.Sidharth,M.P.Sebastian," A Revised Secure Authentication Protocol for IEEE 802.16 (e)", International Conference on Advances in Computer Engineering,2010.
[14] M.Holbal,T,Welzer," An Improved Authentication Protocol Based on One-Way Hash Functions and Diffie-Hellman Key Exchange", International Conference on Availability, Reliability and Security,2009.
[15] F.Leu, Y. Huang, C.Hong Chiu," Improving security levels of IEEE802.16e authentication by Involving Diffie-Hellman PKDS", International Conference on Complex, Intelligent and Software Intensive Systems,2010.
[16] Ergang Liu, Kaizhi Huang and Liang Jin,"the design of trusted access scheme base on identity for wimax network" IEEE computer society (International Workshop on Education Technology and Computer Science),2009

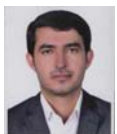
**Mohammad zabihi** recived the B.S. in 1997 and M.S. degree in 2011 in communication engineering from Islamic azad university-shahre rey branch. Now he is an expert in telecommunication company. his research interests include network security, mobile communication and NGN networks.

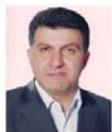
**Ramin Shaghaghi Kandovan** received the B.S. degree in electronic engineering from Tehran University, Tehran, Iran, in 1990 and M.S. and Ph.D. degrees in communication engineering from Islamic Azad University, Tehran, Iran, in 1993 and 2002, respectively. He is currently an Associate Professor in Communication Engineering Department of Islamic Azad University-Shahre Rey Branch. His research interests include Higher Order Statistics, security in communication networks, signal processing. He has been the Head of Post-Graduate Center of IAU-Shahre Rey Branch since 2009.

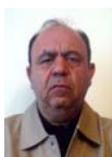
**Mohammad Esmail Kalantari** received the B.S. degree in communication engineering from Communication Technical Faculty, Tehran, Iran in 1972 and M.S. and Ph.D. degrees in communication engineering from Ecole National Superieur des Telecommunications (ENST), Paris, France, in 1979 and 1982, respectively. He has been an Assistant Professor in Electrical Engineering Department of Khaje Nasir Toosi University of Technology for 30 years. Now, he is an academic member of Islamic Azad University, Shahre-Rey Branch, Iran. His research interests include security in communication networks, next generation networks, and mobile communication systems.